\documentclass[conference]{IEEEtran}
\IEEEoverridecommandlockouts
\usepackage{cite}

\usepackage{amsmath}
\usepackage{graphicx}
\usepackage{subcaption}
\usepackage{tabularx}
\usepackage{amsmath}
\usepackage{amssymb}
\usepackage{euler}
\usepackage{bbm}
\usepackage{CJK}
\usepackage{listings}
\usepackage{xcolor}
\usepackage{algorithm}
\usepackage{algpseudocode}
\usepackage{comment}
\usepackage{eurosym}
\def\BibTeX{{\rm B\kern-.05em{\sc i\kern-.025em b}\kern-.08em
    T\kern-.1667em\lower.7ex\hbox{E}\kern-.125emX}}

\usepackage[letterpaper, left=0.625in, right=0.625in, top=0.75in, bottom=1.05in]{geometry}
    
\begin{document}

\title{Pair-Bid Auction Model for Optimized Network Slicing in 5G RAN}
\author{
    \IEEEauthorblockN{Mengyao Li \IEEEauthorrefmark{1},  
        Sebastian Troia\IEEEauthorrefmark{1}, 
        Yingqian Zhang\IEEEauthorrefmark{2}, and 
        Guido Maier\IEEEauthorrefmark{1}
    }
    \IEEEauthorblockA{
        \IEEEauthorrefmark{1}Politecnico di Milano, Milan, Italy 
        \IEEEauthorrefmark{2}TU Eindhoven, Eindhoven, Netherlands
    }
}

\maketitle

\begin{abstract}
Network slicing is a key 5G technology that enables multiple virtual networks to share physical infrastructure, opti- mizing flexibility and resource allocation. This involves Mobile Network Operators (MNO), Mobile Virtual Network Operators (MVNOs), and end users, where MNO leases network slices to MVNOs, and then provides customized services. This work considers end-to-end network slicing with a focus on fair sharing and financial-related power efficiency, modeled as a two-level hierarchical combinatorial auction. At the upper level, an MNO auctions slices to competing MVNOs, while at the lower level, MVNOs allocate resources to end users through their own auctions. Dynamic user requests add complexity to the process. Our model optimizes resource allocation and revenue generation using a pair-bid mechanism and Vickrey-Clarke-Groves (VCG) pricing. The pair-bid approach enhances competition and effi- ciency, while VCG ensures truthful bidding based on marginal system impact. Simulations validate the model's effectiveness in resource distribution and financial performance, showing a 12.5\% revenue improvement over the baseline.
\end{abstract}


\section{Introduction}
\label{sec:intro}

Network slicing partitions physical network resources into virtual networks tailored to specific applications. This technology enhances the 5G RAN capabilities and is expected to unlock further potential in 6G. By enabling efficient resource allocation, isolation, and scalability, network slicing meets diverse service demands such as latency, bandwidth, and reliability.
In 5G RAN, network slices are processed by baseband functions distributed across Centralized Units (CUs), Distributed Units (DUs), and Radio Units (RUs), following the 3GPP functional split \cite{ahmadi2009overview}. CUs and DUs are deployed as Virtual Network Functions (VNFs) in nodes, while RUs remain at antenna sites \cite{9946423}. 5G network slicing has revolutionized network infrastructure, enabling the creation of virtualized, customized network segments tailored to diverse service requirements \cite{10469006}. 
The 5G optical network is considered a good candidate for the x-haul links, that is, fronthaul (RU-DU), midhaul (DU-CU), and backhaul (CU-GW) \cite{ahmadi2009overview}, as shown in Fig.~\ref{fig:model}, represents how the network provides the service for the slice. Network slicing involves multiple stakeholders: MNO manages slices, MVNOs lease and customize them, and users submit bids \cite{zhu2015virtualization}. Given the rising demand for network flexibility and growing energy costs, optimizing power consumption is important.  Consider a scenario where users in a country seek connectivity through Mobile Virtual Network Operators (MVNOs). Each MVNO provides pricing offers, and users choose the one that delivers the most cost-effective service within their budget. The network slices offered differ by type, such as enhanced Mobile Broadband (eMBB) or ultra-Reliable Low Latency Communications (uRLLC), to accommodate diverse service requirements. Importantly, users rent these slices for a limited duration, rather than occupying them indefinitely, allowing for dynamic allocation and resource efficiency over time.

This work presents a two-level hierarchical combinatorial auction model for 5G RAN network slicing \cite{9411723}.
Upper-Level Auction (single seller, multiple buyers): The MNO auctions slices to MVNOs, maximizing profit while considering electricity costs and fair resource distribution.
Lower-Level Auction (multiple sellers, multiple buyers): MVNOs lease acquired slices to users, who bid dynamically across multiple MVNOs. MVNOs optimize allocation and pricing based on resale gains to enhance revenue.
A key innovation of our model is its ability to handle dynamic user requests, allowing them to enter and exit at different timeslots—an approach that closely mirrors real-world network operations. Additionally, the pair-bid mechanism fosters competition by enabling users to express their MVNO preferences while allowing MVNOs to dynamically adjust pricing strategies for optimal resource allocation. This ensures an efficient matching process between users and MVNOs, as detailed in Section \ref{sec:hierarchical}.

This study tackles two key challenges in network slicing. First, it introduces a financial perspective that accounts for monetary costs and revenue generation. Our model integrates power consumption into the economic model, incorporating a power-efficient CU-DU placement strategy \cite{zorello2022power} to reduce system costs and energy usage. At the upper level, it focuses on allocating network slices to MVNOs to maximize MNO revenue while minimizing electricity expenses, building on our previous hierarchical auction model \cite{zorello2023auction}. Second, the lower-level auction manages dynamic user requests using a multiple-seller, multiple-buyer structure and a pair-bid mechanism to simulate competition among MVNOs. The goal is to maximize the margin between MVNO bids and user bids. The model enforces end-to-end constraints, including latency, throughput, and baseband functional split requirements. 

To ensure truthful bidding, we adopt the Vickrey-Clarke-Groves (VCG) pricing mechanism \cite{vickrey1961counterspeculation}. We formulate the resource allocation problem as a Winner Determination Problem (WDP) \cite{zhu2015virtualization}, aiming to maximize overall revenue. To solve it under dynamic demand, we design a heuristic algorithm that accommodates time-varying requests while preserving the integrity of the pair-bid mechanism. To the best of our knowledge, this is the first network slicing framework that incorporates a financial cost model with dynamic auctions, representing a novel contribution to both theory and practice.


The remainder of this paper is structured as follows. Sec.~\ref{sec:related_works} presents a discussion of related works. Sec.~\ref{sec:hierarchical} defines the problem statement and outlines the ILP model. Sec.~\ref{sec:solve} introduces our novel and scalable heuristic algorithm. Sec.~\ref{sec:results} provides a detailed numerical analysis. Finally, Sec.~\ref{sec:conclusion} summarizes this study.

\section{Related Works}
\label{sec:related_works}

This section reviews related work on network slicing, focusing on 5G RAN applications and auction-based resource allocation. The 5G RAN model supports various services, including uRLLC and eMBB \cite{ahmadi2009overview}, with power consumption influenced by x-haul links and radio unit placement \cite{zorello2022power}.

Zhu et al.\cite{zhu2015virtualization} explored wireless virtualization, proposing a hierarchical combinatorial auction to allocate resources efficiently in large networks.
Zheng et al.\cite{9564520} introduced a hierarchical RAN slicing model, facilitating slice resource sharing through VCG pricing to optimize transactions.
Peng et al.\cite{peng2024intelligent} developed a hierarchical auction framework using Particle Swarm Optimization (PSO) to improve spectrum allocation for MVNOs and power terminals.
Xiang et al.\cite{8198811} designed a Fog RAN slicing framework ensuring QoS across latency, energy, and reliability requirements.
Sun et al.\cite{sun2020paired} proposed a pair-bid double-auction model, where MVNOs and MNOs submit bids to a Network Slice Broker (NSB) for social welfare optimization.
Promponas et al.\cite{10228856} presented an iterative market model using double-sided auctions for multi-MNO network slicing.
Jiang et al.\cite{jiang2017network} proposed a business network model for delivering 5G network slices, ensuring optimal allocation of computational and storage resources to meet service requirements.

\textbf{Novelty:} We propose a resource allocation method that not only satisfies service-specific requirements but also prioritizes minimizing overall network power consumption. A key innovation and novelty of our model is its financial perspective—unlike traditional approaches  \cite{zhu2015virtualization, 9564520, peng2024intelligent, 8198811, sun2020paired, 10228856, jiang2017network} that primarily focus on resource utilization, traffic, QoS, our framework explicitly incorporates real-world monetary costs, including electricity expenses and bid-related payments. In our formulation, resource usage, traffic, latency, QoS are treated as constraints, while the main objective is to minimize total financial expenditure.
Our work introduces a two-level auction model involving three distinct types of entities and integrates a pair-bid mechanism \cite{sun2020paired, 10228856}. This model builds on our previous research \cite{zorello2023auction}, which serves as the baseline for comparison. We also use \cite{jiang2017network} as the second baseline. Unlike the baseline, which does not support dynamic requests, multi-buyer and multi-seller auctions, or pair-bid mechanisms, our heuristic optimizes revenue by maximizing the difference between MVNO bids (cost estimation and resale gain) and user bids. In contrast, the baseline focuses only on maximizing the number of accepted user bids, limiting its flexibility.
A key improvement in our model is that users are allowed to dynamically choose between different MVNOs based on price and service quality, resulting in a more realistic and competitive multi-buyer, multi-seller network environment.

\section{Two-level Hierarchical model }
\label{sec:hierarchical}
\begin{figure}[b]
\vspace{-6mm}
    \centering \includegraphics[width=0.45\textwidth]{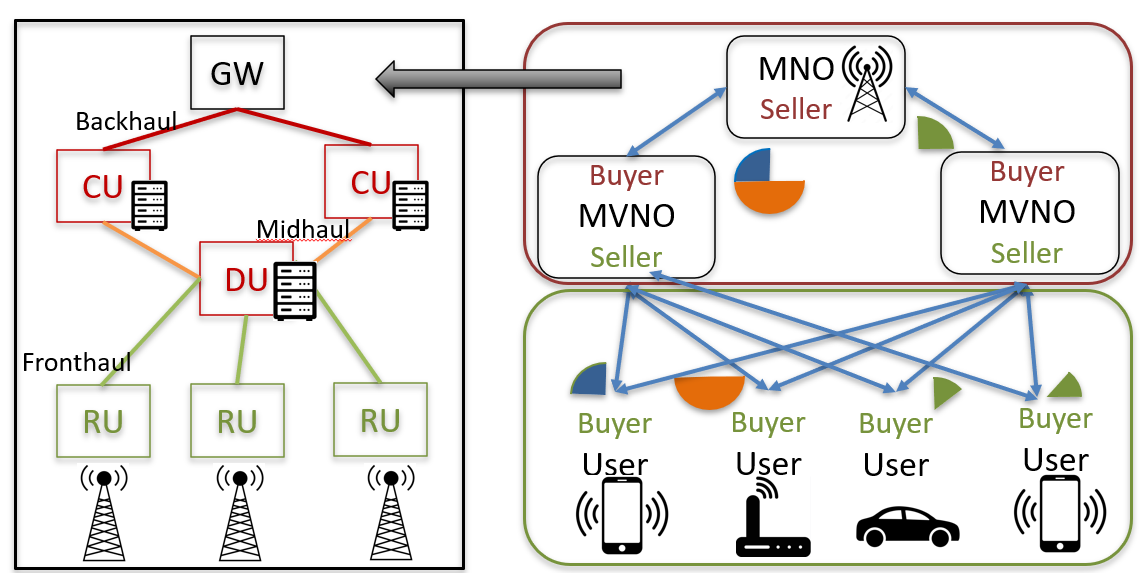}
    \caption{Hierarchical auction model build on 5G RAN network.}
    \label{fig:model}
\end{figure}

Fig. \ref{fig:model} illustrates the hierarchical auction model for network slicing, incorporating network baseband functions (CU, DU, RU, and core GW). The model operates on two auction levels:
Upper-Level Auction: A single-seller, multiple-buyer auction where the MNO acts as the seller, offering slices to MVNOs. MVNOs bid on sets of slices based on origin node, service delay, and capacity. The MNO assigns service delays per explicit requests, offering slices of varying sizes and colors (Fig. \ref{fig:model}).
Lower-Level Auction: A multiple-seller, multiple-buyer auction where MVNOs resell and subdivide slices to users, who bid across multiple MVNOs. Unlike users, MVNOs focus on slice allocation and resale gain rather than intrinsic service requests.
For example, as shown in Fig. \ref{fig:model}, an MVNO may purchase a green slice, verify its origin, delay, and capacity, and then subdivide it for different users, ensuring efficient utilization. This two-level structure allows users to select MVNOs, with transient requests having defined start and end times. 
A key feature of the lower-level auction is the pair-bid mechanism, where MVNOs and users bid for each other, distinguishing our model from static approaches. Fig. \ref{fig:model} shows blue arrows for the possibility of MVNOs and users bidding for each other. Users dynamically evaluate MVNOs, while MVNOs assess costs and resale gains before bidding, making pair-bid auctions an essential part of the model \cite{sun2020paired}.

\begin{table}
\footnotesize
\caption{Decision variables and parameters for lower level}
\begin{tabularx}{\linewidth}{p{0.7cm} X}

\hline
\textbf{Sets}& \textbf{Description} \\
\hline \hline

$K$ & Set of users' requests\\

$U$ & Set of radio network units ${RU, CU, DU, GW}$\\

$T$ & Set of timeslot\\
$N$ & Set of nodes\\
$V$ & Set of MVNOs\\
$S$ & Set of slices\\

\hline
\textbf{Para} & \textbf{Description} \\
\hline \hline
$b^k_{p,q}$ & Integer, user request $k$ initiates at timeslot $p$ and concludes at timeslot $q$\\
$\lambda_k$ & Integer, traffic demand for request $k$\\
$N_s$ & Integer, the original node of slice $s \in S$\\
$n_k$  & Integer, the original node of request $k \in K$\\
$T_s$ & Integer, service type associated with slice $s \in S$\\
$t_k$ & Integer, service types needed by request $k$\\
$C_s$ & specifies the maximum capacity of slice $s \in S$\\
$L_v^s$ & represents the upper capacity bound for MVNO $v \in V$\\
\hline
\textbf{Var} & \textbf{Description} \\
\hline \hline
$\beta^t_{v,k}$ & Integer, the bid proposed by MVNO $v$ to the user request $k \in  K$\\
$d_k^t$ & Binary, whether the MVNO accepts request $k \in K$ at timeslot $t \in T$\\
$r_{v,k}^{s,t}$ & Binary, determines whether request $k \in K$ is allocated to slice $s \in S$ belonging to MVNO $v \in V$\\
\hline
\end{tabularx}
 \centering
\vspace{-4mm}
\end{table}

\subsection{Formulas for Lower-level Auction}
\label{sec:ll_auction}

In our auction model, each MVNO-user slice transaction must form a unique pairing. In a double auction framework, revenue maximization is achieved by allocating slices with the lowest MVNO bid to users with the highest bids. This strategy remains effective provided that all participants truthfully disclose their valuations. To determine winners, we apply natural ordering, matching the highest-bidding user with the lowest-bidding MVNO, thereby ensuring both efficiency and fairness in slice allocation. Additionally, each request is associated with an arrival timeslot $p$ and a departure timeslot $q$, introducing a dynamic flow of requests into the system. Eq.~\ref{ll_obj} maximizes the sum value difference between the bids from MVNOs and users. 
Eq.~\eqref{ll_SingleSlice} ensures that request $k\in K$ (if accepted) is admitted into a single network slice $s \in S$ owned by the MVNO $v$. 
Eq.~\eqref{ll_OriginType} ensures origin node and service type match between user request $k\in K$ and network slice $s\in S$. 
Eq.~\eqref{ll_SliceCapacity} limits total traffic carried by the slice $s\in S$ to the slice capacity $C_s$.
Eq.~\eqref{ll_MVNOCapacity} limits total traffic carried by all the slice $s\in S$ related to MVNO $v$ lower than upper bound capacity $L_v^s$.

\footnotesize
\begin{equation}
\label{ll_obj}
\max \sum_{k \in K, v \in V, t \in T} (b^k_{p,q} - \beta^t_{v,k}) d_{v,k}^t \qquad 
\end{equation}
\footnotesize
\begin{equation}
\label{ll_SingleSlice}
\sum_{s \in S} r_{v,k}^{s,t} = d_{v,k}^t \qquad \forall k\in K, t\in T
\end{equation}
\footnotesize
\begin{equation}
\label{ll_OriginType}
r_{v,k}^{s,t} \leq 
   \begin{cases}
      d_{v,k}^t, & \text{if } N_s = n_k \text{ and } T_s = t_k, \\
      0, & \text{otherwise}.
    \end{cases} \forall u \in U, \ \forall k \in K
\end{equation}
\footnotesize
\begin{equation}
\label{ll_SliceCapacity}
\sum_{k \in K} r_{v,k}^{s,t} \lambda_k \leq C_s \qquad \forall s \in S
\end{equation}
\normalsize

\footnotesize
\begin{equation}
\label{ll_MVNOCapacity}
\sum_{k \in K} \sum_{s \in S} r_{v,k}^{s,t} \lambda_k \leq L_v^s \qquad \forall v \in V
\end{equation}

\normalsize

\normalsize

\subsection{Formula for upper-level auction}
\label{sec:ul_auc}
\normalsize
We represent the 5G RAN architecture using a graph structure, denoted as $G$, where the set of nodes is $N$ and the set of physical links is $E$. The MVNO must guarantee that traffic within each slice is transmitted from the source $RU$ to the $GW$, ensuring it is processed through a sequence of radio-network components defined as $U = {RU, DU, CU, GW}$. 

\footnotesize
\begin{equation}
\label{UL_obj}
\max \sum_{j \in J, t \in T} b^j_t \cdot f^j_t - (P_{net} + P_{node}) EC
\end{equation}
\normalsize \normalsize
The objective function in the upper-level is to maximize the MNO's profit, as defined in Eqn~\eqref{UL_obj}, which represents the difference between revenue and electricity expenses. $b^j_t$ means bid value from the MVNO request $j \in J$ in the upper level, while $f^j_t$ represents if the MVNO request $j \in J$ was accepted by the
MNO. The revenue component is determined by summing the total value of accepted bids, while the cost is derived from the overall power consumption multiplied by the electricity cost ($EC$).
The power consumption of network infrastructure is represented by $P_{net}$, which includes the energy usage of the front-haul, mid-haul, and back-haul interfaces. Additionally, $P_{node}$ accounts for the total power consumption of network nodes. To determine the deployment of CU and DU, and decrease the power consumption, we reference the model from our previous research Ref. \cite{zorello2022power} Eqns. 1-21, which assesses actual power consumption, electricity expenditure, and the most efficient configuration of radio units, capacity constraints, traffic constraints, split latency requirements, QoS maintenance constraints.

\footnotesize
\begin{equation}
\label{mvno_value}
\textit{bid value} = (R_1 - R_2)*(1 - \textit{resale gain})
\end{equation}
\normalsize
$R_1$ represents the revenue considering both current and newly acquired slices, while $R_2$ accounts for revenue from only the currently held slices. The \textit{resale gain} is a fixed percentage defined by each MVNO, indicating the portion of user payment retained as profit.
As shown in Eqn.~\eqref{mvno_value}, an MVNO's bid is determined solely by its user demand and potential resale gains. Rather than acting as single-minded bidders, MVNOs treat the upper- and lower-level auctions as separate processes, prioritizing resource acquisition over strategic interdependence. To increase their probability of securing resources, they submit bids for multiple capacity variations of the same slice.
To ensure that only one bid per slice type and capacity is selected, we implement XOR bidding. Once bids are collected, the MNO executes the upper-level WDP, selects winning bids, provisions slices, assigns DUs and CUs, and establishes traffic routing.

\textbf{VCG Pricing}: The formulated WDPs ensure fair resource sharing while promoting truthfulness, meaning bidders' optimal strategy is to bid their true valuation. The VCG auction, a widely used offline auction mechanism, guarantees truthfulness and is effective for network resource allocation \cite{vickrey1961counterspeculation, talwar2003price}. As a generalization of the second-price auction for multiple commodities, VCG preserves incentive compatibility by charging bidders based on the potential loss they impose on others.
Truthfulness in VCG holds when the auction mechanism satisfies monotonicity. Monotonicity ensures that a bidder’s chances of winning a resource bundle increase by either raising their bid or reducing the requested resources \cite{vickrey1961counterspeculation, zhu2015virtualization, sun2020paired}.  Specifically, a bidder can improve their ranking by increasing their bid value or lowering their demand, reinforcing truthful bidding.
Thus, under VCG pricing, the bid is defined:

\footnotesize
\begin{equation}
\label{vcg}
b^{VCG,k}_t = \sum_{j\neq k} b^j_{p,q} (\overline{S}_j^*) - \sum_{j\neq k} b^j_{p,q}(S_j^*)   \qquad p \leq t \leq q
\end{equation}

\normalsize 
In this model, $\overline{S}_j^*$ and $S_j^*$ denote the resources allocated to bidder $j \in J$ \cite{zhu2015virtualization}. Since the network aims to maximize social welfare, the VCG mechanism charges winners based on their social cost. This cost is determined as the difference between the maximum feasible welfare without bidder $j$ and the welfare to others when $j$ is included \cite{zhu2015virtualization}.
While VCG pricing ensures truthfulness and fairness, it does not necessarily maximize the seller’s revenue. When resources are abundant and competition is low, the VCG price may drop to zero, yielding no revenue. To mitigate this, a base access price is introduced as the minimum amount a bidder must pay for network slices \cite{zhu2015virtualization}. Each slice type is assigned a predefined base access price. The final charge for a bidder is:
$q_k = \max \{q_k^{Base}, q_k^{VCG}\}$.
The seller sets and announces these base prices beforehand. This hybrid approach ensures that when resources are plentiful, the base access price guarantees revenue, and when resources are scarce, the VCG price promotes fair allocation and truthful bidding. Through both mechanisms, the auction maintains truthfulness while securing a minimum revenue for the seller.

\begin{figure}
    \centering \includegraphics[width=0.4\textwidth]{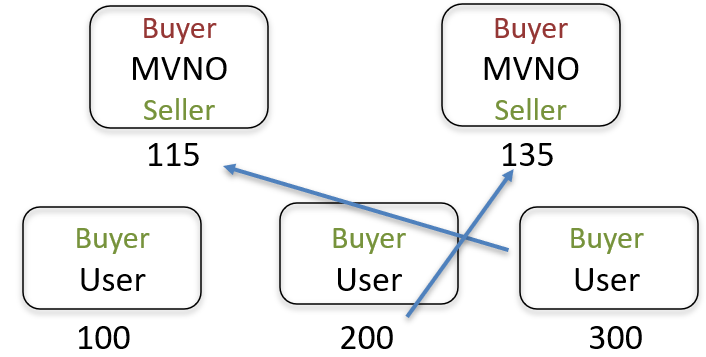}
    \caption{An example to understand pair-bid.}
    \label{fig:ex}
    \vspace{-6mm}
\end{figure}

In Fig.~\ref{fig:ex}, two MVNOs each offer a single network slice with a capacity of 1 Gbps, providing a total of two available slices. Users 1, 2, and 3 submit bids of 100, 200, and 300, respectively, competing for the limited slices, meaning one user will not secure a slice.
As this auction follows a pair-bid mechanism, MVNOs must also submit price bids for users. MVNO1 offers a price of 115 to all users, while MVNO2 sets its price at 135. The VCG base price is 135, and each MVNO can only accommodate one request due to capacity constraints. Given the highest bid of 300, User 3 has priority and selects MVNO1, which offers the lower price of 115. User 2, with the second-highest bid of 200, is then left to choose MVNO2 at the price of 135, while User 1 does not secure a slice.
Following the VCG, the final bid value is adjusted to the base price of 135, ensuring fairness and revenue stability.

\section{Algorithm for Hierarchical Auction Model} 
\label{sec:solve}
Both the lower-level and upper-level WDPs are formulated as combinatorial auction models, utilizing Integer Linear Programming (ILP) to determine the optimal solution. In the lower level, the WDP involves multiple users interacting with multiple MVNOs, whereas at the upper level, a single MNO engages with multiple MVNOs. Prior research has established that this class of problems is NP-Hard \cite{zhu2015virtualization}. Furthermore, our previous work \cite{zorello2022power} demonstrated that the placement of DUs and CUs, along with routing decisions, also falls into the NP-Hard category.
To solve these models, we designed heuristic algorithms, building upon our prior research in \cite{zorello2023auction}.

The core requirement for VCG truthfulness is monotonicity—where a bidder’s chances of winning increase by raising their bid or reducing their resource demand. Greedy algorithms naturally satisfy this through ranking-based allocation. Bidders are sorted by metrics like bid-per-resource, and resources are assigned iteratively to the top-ranked ones. Raising a bid or requesting less improves a bidder’s rank and feasibility, ensuring monotonicity. As a result, the greedy approach supports truthful bidding, aligning with the VCG pricing mechanism.

\subsection{Lower-level auction algorithm}

The lower-level auction mechanism is shown in Algorithm~\ref{alg:ll_greedy}. To uphold the monotonicity of user-provided bids and maintain truthfulness, the set of requests $K$ is arranged in descending order according to their origin node $n_k$ and requested type of service $t_k$. This ordering is based on a normalized metric \cite{zhu2015virtualization} to ensure truthful bidding: $\frac{b_k}{\sqrt{\lambda_k}}$, where $b_k$ signifies the bid amount, and $\lambda_k$ denotes the traffic demand for request $k$. This prioritization strategy favors requests with higher bid-to-traffic ratios, ensuring that network slices are distributed to those with the most substantial relative demand. Network slices $S$ are also sorted in order, according to their capacity $C_s$.  By assigning resources to bids with higher priority, fairness is maintained as long as the pricing structure preserves bidder truthfulness, as outlined in Section \ref{sec:ll_auction}. The algorithm then proceeds to match MVNO bids to users, and check for the upper bound of slices. A pair-bid approach is employed, ensuring that the most cost-efficient goods are assigned to buyers placing the highest bids. Finally, for every combination of (origin node and service type), resources are allocated to the top request in the ordered list, facilitating an optimized and prioritized allocation process.

\begin{algorithm}
\footnotesize
  \caption{Greedy Algorithm for the lower-level WDP}
  \label{alg:ll_greedy}
  \begin{algorithmic}[1]
    \Require Slices $S$, requests $K$
    \Ensure Slice allocation, fair sharing
    \State Initialize occupied capacity $O_s = 0$ for each slice $s\in S$
    \State Classify submitted requests $K$ based on their origin node $n_k$ and requested service $t_k$ for each timeslot $t$
    \State Classify slices $S$ based on their origin node $N_s$ and requested service $T_s$
    \For{each possible proposal set of request and bid $l\in L$}
        \State Calculate $\frac{b_k}{\sqrt{\lambda_k}}$ for each request $k\in K_l$
        \State Sort and re-index the requests $K_l$: $\frac{b_1}{\sqrt{\lambda_1}} \geq \frac{b_2}{\sqrt{\lambda_2}} \geq ... \geq \frac{b_k}{\sqrt{\lambda_k}}$
        \State Sort and re-index the slices $S_l$: $C_1 \leq C_2 \leq ... \leq C_s$
        \For{each request $k\in sorted(K_l)$}
            \For{each slice $s\in sorted(S_l)$}
                \State Find the pair for each MVNO and User acorrding to the bids
                \If{$\lambda_k + O_s \leq C_s$}
                    \State $a_k = 1$, $z_s^k = 1$, $O_s = O_s + \lambda_k$
                \EndIf
            \EndFor
        \EndFor
    \EndFor
    \State Call Algorithm 2 and get slices from MNO
    \State Process user requests $k$ in descending order and evaluate whether each can be allocated a slice from the corresponding MVNO
    \State Calculate the final revenue based on the accepted request
  \end{algorithmic}
\normalsize
\end{algorithm}
\normalsize
\vspace{-4mm}

\subsection{Upper-level auction algorithm}

At the upper level of our model, the WDP is addressed using a greedy strategy. 
Algorithm~\ref{alg:ul_greedy} outlines the greedy solution for the upper-level WDP. For each timeslot $t$, all bids $J$ from MVNOs are ranked in descending order using a normalized benefit metric $NB_j$, which helps maintain monotonicity—a key condition for supporting truthful behavior under VCG pricing.
This normalized benefit is calculated from bid value $b_j$, number of requested slices $r_j$, slice set $I_j$ for bid $j$, with $\lambda_i^u$ representing traffic demand for each slice $i \in I_j$, and $t_i^{max}$ denoting the maximum latency requirement for the same slice. The ranking approach ensures that bids are sorted by value, bandwidth requirements, and latency expectations in a manner that upholds monotonicity.

\vspace{-2mm}
\begin{algorithm}[hbpt]
\footnotesize
  \caption{Greedy Algorithm the upper-level WDP}
  \label{alg:ul_greedy}
  \begin{algorithmic}[1]
    \Require Network Graph $G(N,E)$, MVNOs' bids $J$
    \Ensure Slice, DU/CU placement, routing, fair sharing
    \State Calculate $NB_j$ for each bid $j\in J$ at timeslot $t$
    \State Sort and re-index the bids $J$ for each $t$: $NB_1 \geq NB_2 \geq ... \geq NB_j$
    \For{each bid $j\in sorted(J)$}
        \State $B = B + I_j$
        \If{placement(B) is feasible}
            \State $f_j = 1$,$D = B$
        \Else
            \State $B = B - I_j$
        \EndIf
    \EndFor
  \end{algorithmic}
\end{algorithm}
\vspace{-2mm}

\normalsize

Following this sorting, the algorithm iteratively constructs a candidate selection set $B$. For every bid $j \in J$, it is provisionally added to $B$, and the system checks whether CU and DU placement along with routing can be supported. This feasibility check leverages a heuristic adapted from \cite{zorello2022power}, designed to consider power usage, latency, and infrastructure constraints.
If the system can accommodate the placement, the bid is officially added to the final accepted set $D$; if not, it is discarded from $B$. This mechanism ensures only feasible bids are selected, improving efficiency without violating system limitations. The overall time complexity of the greedy algorithm is $O(n\log n)$, where $n$ is the number of MVNO bids, making it scalable for dynamic and real-time resource allocation in RAN slicing.

\footnotesize
\vspace{-2mm}
\begin{equation}
\label{ul_sort}
NB_j = \frac{b_j}{\sqrt{w_1 \frac{\sum_{i\in I_j} \lambda_i^{u}}{r_j} + w_2 \left(\frac{\sum_{i\in I_j} t_i^{max}}{r_j}\right)^{-1}}} \qquad \forall j\in J
\end{equation}
\normalsize

\section{results} \label{sec:results}
This section evaluates the auction performance by comparing our heuristic algorithm against the baseline. We assess the heuristic’s performance in a dynamic request scenario to further validate its adaptability and efficiency.

\subsection{Simulation settings}
For evaluation, we adopt the network topology from \cite{zorello2022power}. Each node is equipped with 8 cores, operates at 3.7 GHz, and has a processing capacity of 537.6 GFLOPS. Power consumption varies between 130 W in idle mode and 870 W under full load. The fronthaul, midhaul, and backhaul interfaces require 18.2 W, 10 W, and 1 W, respectively \cite{askari2020dynamic}. RU deployment follows the industry standards outlined in \cite{small2016small}. Nodes are interconnected through bidirectional fiber links with a capacity of 100 Gbit/s, utilizing 100 Gbit/s transponders that consume 110.4 W \cite{elmirghani2018greentouch}.

\begin{figure}
    \centering \includegraphics[width=0.46\textwidth]{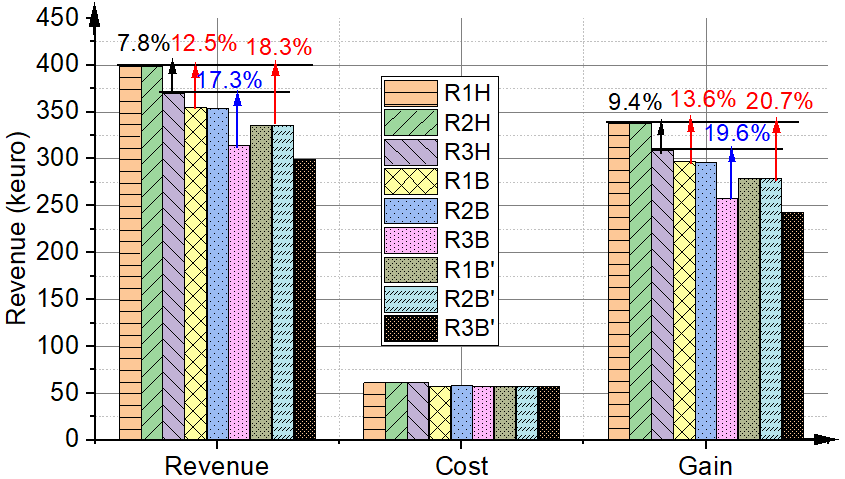}
    \caption{Result for MNO revenue}
    \label{fig:MNO1}
\vspace{-4mm}
\end{figure}

To benchmark our approach, we compare it against the baseline1 method from \cite{zorello2023auction}, which does not incorporate dynamic request handling, multi-buyer/seller auction mechanisms, or pair-bid strategies. And another baseline2 algorithm from \cite{jiang2017network} which performs request admission without any sorting mechanism and users are single-mind to a MVNO. In contrast, our heuristic is designed to maximize the gap between MVNO bids—factoring in both cost and resale gain—and user bids, thereby enhancing revenue optimization. Moreover, our model introduces a more realistic market dynamic by allowing users to select among multiple MVNOs, improving flexibility in resource allocation.

\subsection{Result comparison}

\begin{figure}
    \centering \includegraphics[width=0.46\textwidth]{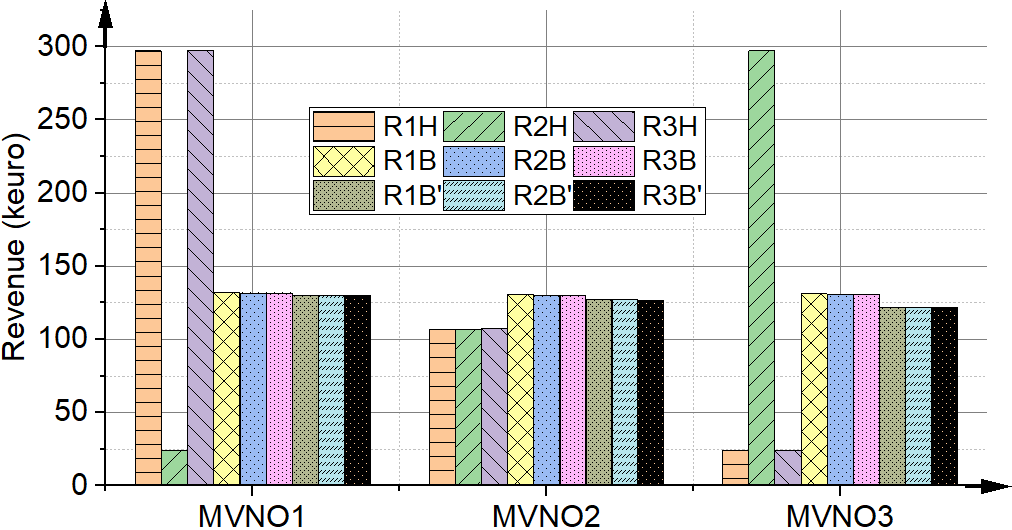}
    \caption{Result for three MVNO revenue}
    \label{fig:MVNO1}
 \vspace{-6mm}
\end{figure}

To evaluate our heuristic algorithm, we test a 10-node topology under diverse settings. Experiments involve three MVNOs in a hierarchical auction framework, handling user requests with capacities uniformly distributed (0.5–5 Gbps) and an equal number of requests per MVNO. We also examine cases where all MVNOs have fixed 5 Gbps slice capacities.
We evaluate MVNO revenue under three resale gain configurations: R1: [5\%, 10\%, 15\%], R2: [15\%, 10\%, 5\%], and R3: [10\%, 20\%, 30\%], where each value represents the resale gain assigned to one of the three MVNOs.
The heuristic algorithm (H) is compared with the baseline1 (B) from \cite{zorello2023auction}, and baseline2 (B') from \cite{jiang2017network}, where "R1H" represents the heuristic applied to R1. We test 10 timeslots with an equivalent number of requests per timeslot to assess stability and efficiency.
The objective functions differ: the heuristic maximizes the difference between user and MVNO bids, optimizing profit, while the baseline maximizes total accepted user bids. For fairness, when we compare the baseline with the heuristic, all results use the sum of accepted user bids as the evaluation metric. Despite this, our heuristic consistently outperforms the baseline in revenue, as shown in the results. $Gain = Revenue - Cost$



Next, we analyze performance gaps using visual indicators. Black arrows mark differences between heuristic algorithms (R1 and R3), red highlights gaps between heuristic and baseline in R1 and R2, and blue represents the gap in R3.
Fig.~\ref{fig:MNO1} illustrates MNO revenue, cost, and profit across both approaches. The heuristic algorithm maintains stable revenue for R1 and R2, while R3 experiences a 7.8\% drop due to higher MVNO resale gains.
Compared to the heuristic, the baseline1 generates 12.5\% lower revenue for R1 and R2, with a 17.3\% gap for R3. The baseline2 generates 18.3\% lower revenue for R1 and R2, with a 23\% gap for R3. The heuristic incurs 6.9\% higher costs, as it accepts more requests. These visualized gaps highlight the heuristic’s superior performance and resource efficiency.

\vspace{-3mm}
\begin{figure}[hbpt]
    \centering \includegraphics[width=0.45\textwidth]{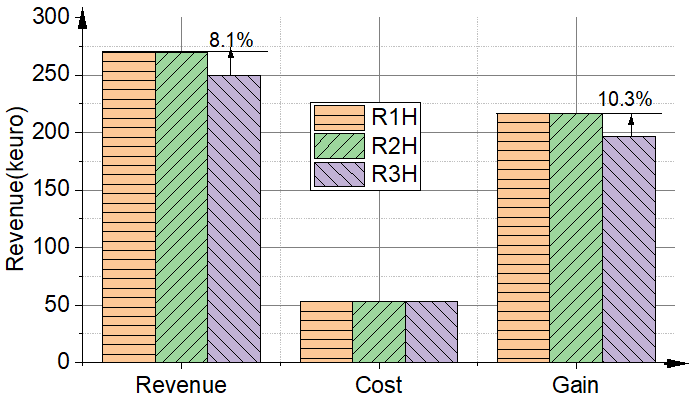}
    \caption{Result for MNO revenue with dynamic requests.}
    \label{fig:MNO2}
    \vspace{-8mm}
\end{figure}

\begin{figure}[hbpt]
    \centering \includegraphics[width=0.45\textwidth]{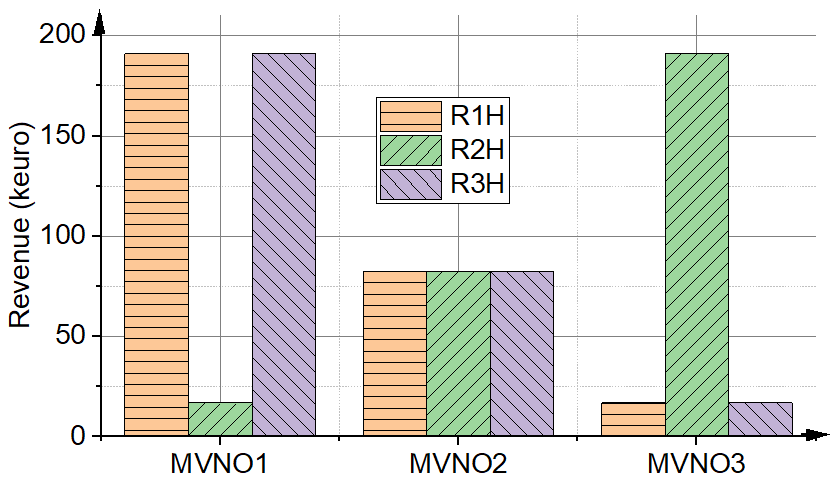}
    \caption{Result for MVNO revenue with dynamic requests.}
    \label{fig:MVNO2}
    \vspace{-3mm}
\end{figure}

Fig.~\ref{fig:MVNO1} illustrates MVNO revenue under different resale gains. In the baseline1 and baseline2, MVNO revenue remains nearly constant since users bid exclusively for one MVNO, eliminating competition. In contrast, our heuristic fosters competition, leading to differentiated revenues. In R1 and R3, MVNO1 (lowest resale gain) earns the most, while MVNO3 (highest resale gain) earns the least. In R2, the trend reverses. Overall, the heuristic improves total MVNO revenue by 8.6\% over the baseline1, and 12\% over baseline2.
Next, Fig.\ref{fig:MNO2} evaluates MNO revenue under dynamic requests scenario, making direct baseline comparison unsuitable. R1 and R2 achieve the same revenue, while R3 sees an 8.1\% drop due to higher resale gains. Despite revenue variations, costs remain stable, indicating consistent request acceptance. Overall revenue is lower than in Fig.\ref{fig:MNO1} because dynamic requests exit the system earlier, reducing processed requests.
Finally, let's compare the MVNO revenue under varying resale gains shown in Fig. \ref{fig:MVNO2}. 
A similar trend to Fig. \ref{fig:MVNO1} is observed, where the scenarios R1 and R2 exhibit reversed results due to differences in resale gains. Despite these variations, the total revenue across all MVNOs remain consistent, demonstrating the stability of the model.

\section{Final remarks}
\label{sec:conclusion}
We propose a two-level hierarchical auction framework for network slicing in 5G RAN, addressing the challenges of dynamic resource allocation and revenue optimization in competitive environments. Simulation results show that our model improves revenue by 12.5\% over baseline 1 and 18.3\% over baseline 2, while also enhancing resource utilization and maintaining QoS guarantees. This framework offers a practical, scalable solution for efficient resource management and service delivery in next-generation wireless networks. As future work, we aim to explore larger network topologies and further refine the upper-level auction mechanism to enhance efficiency in subsequent studies.

\section*{Acknowledgment}
\footnotesize
This work was supported by the European Union - Next Generation EU under the Italian National Recovery and Resilience Plan (NRRP), Mission 4, Component 2, Investment 1.3, CUP D43C22003080001, partnership on “Telecommunications of the Future” (PE00000001 - program “RESTART”).

\bibliographystyle{ieeetr}
{\footnotesize
\bibliography{sample}}

\begin{thebibliography}{10}

\bibitem{ahmadi2009overview}
S.~Ahmadi, ``An overview of 3gpp long-term evolution radio access network,'' {\em New Directions in Wireless Communications Research}, 2009.

\bibitem{9946423}
R.~Joda~\textit{et al.}, ``Deep reinforcement learning-based joint user association and cu–du placement in o-ran,'' {\em IEEE TNSM}, 2022.

\bibitem{10469006}
A.~Sukumar~\textit{et al.}, ``Enhancing security and privacy implications in 5g network slicing,'' in {\em 2024 ICAECT}, pp.~1--8, 2024.

\bibitem{zhu2015virtualization}
K.~Zhu~\textit{et al.}, ``Virtualization of 5g cellular networks as a hierarchical combinatorial auction,'' {\em IEEE TMC}, 2015.

\bibitem{9411723}
S.~D’Oro~\textit{et al.}, ``Coordinated 5g network slicing: How constructive interference can boost network throughput,'' {\em IEEE/ACM TON}, 2021.

\bibitem{zorello2022power}
L.~M.~M. Zorello~\textit{et al.}, ``Power-efficient baseband-function placement in latency-constrained 5g metro access,'' {\em IEEE TGCN}, 2022.

\bibitem{zorello2023auction}
L.~M.~M. Zorello~\textit{et al.}, ``Auction-based network slicing for 5g ran,'' in {\em 2023 IEEE NetSoft}, pp.~390--395, IEEE, 2023.

\bibitem{vickrey1961counterspeculation}
W.~Vickrey, ``Counterspeculation, auctions, and competitive sealed tenders,'' {\em The Journal of finance}, vol.~16, no.~1, pp.~8--37, 1961.

\bibitem{9564520}
W.~Zheng~\textit{et al.}, ``Hierarchical resource allocation for ran slicing in power wireless private network,'' in {\em 2021 IEEE ICSPCC}, 2021.

\bibitem{peng2024intelligent}
Y.~Peng~\textit{et al.}, ``An intelligent resource allocation strategy with slicing and auction for private edge cloud systems,'' {\em Future Generation Computer Systems}, vol.~160, pp.~879--889, 2024.

\bibitem{8198811}
H.~Xiang~\textit{et al.}, ``Network slicing in fog radio access networks: Issues and challenges,'' {\em IEEE Commun. Mag.}, 2017.

\bibitem{sun2020paired}
S.~Sun~\textit{et al.}, ``Paired bid-based double auction mechanism for ran slicing in 5g-and-beyond system,'' in {\em 2020 IEEE ICCT}, IEEE, 2020.

\bibitem{10228856}
P.~Promponas~\textit{et al.}, ``Network slicing: Market mechanism and competitive equilibria,'' in {\em IEEE INFOCOM 2023}, pp.~1--10, 2023.

\bibitem{jiang2017network}
M.~Jiang~\textit{et al.}, ``Network slicing in 5g: An auction-based model,'' in {\em 2017 IEEE ICC}, pp.~1--6, IEEE, 2017.

\bibitem{talwar2003price}
K.~Talwar, ``The price of truth: Frugality in truthful mechanisms,'' in {\em STACS}, Springer, 2003.

\bibitem{askari2020dynamic}
L.~Askari~\textit{et al.}, ``Dynamic du/cu placement for 3-layer c-rans in optical metro-access networks,'' in {\em 2020 ICTON}, IEEE, 2020.

\bibitem{small2016small}
S.~C. Forum, ``Small cell virtualization functional splits and use cases,'' 2016.

\bibitem{elmirghani2018greentouch}
J.~M. Elmirghani~\textit{et al.}, ``Greentouch greenmeter core network energy-efficiency improvement measures and optimization,'' {\em JOCN}, 2018.

\end{thebibliography}

\end{document}